# Mathematical Constraint on Functions with Continuous Second Partial Derivatives

J.D. Franson
*Physics Department, University of Maryland, Baltimore County, Baltimore, MD 21250*

**Abstract**
A new integral identity for functions with continuous second partial derivatives is derived. It is shown that the value of any function $f(\mathbf{r},t)$ at position $\mathbf{r}$ and time t is completely determined by its previous values at all other locations $\mathbf{r'}$ and retarded times $t' \leq t$, provided that the function vanishes at infinity and has continuous second partial derivatives. Functions of this kind occur in many areas of physics and it seems somewhat surprising that they are constrained in this way.

## 1. Introduction

Functions with continuous second partial derivatives occur in many areas of physics, such as classical electromagnetism, acoustics, and quantum mechanics. It is shown here that the value of any function $f(\mathbf{r},t)$ at position $\mathbf{r}$ and time t is completely determined by its previous values at all other locations $\mathbf{r'}$ and retarded times $t' \leq t$, provided that the function has continuous second partial derivatives and that it vanishes at infinity. This constraint corresponds to an integral identity that can be derived using Green's function techniques. It seems somewhat surprising that functions of this kind are constrained in this way independent of any assumptions regarding the dynamics of the system. Aside from its fundamental interest, integral identities are often useful in practical applications as well.

The integral identity of interest here is similar in some respects to Cauchy's theorem for analytic functions [1]:

$$f(z) = \frac{1}{2\pi i} \oint_C \frac{f(\xi)d\xi}{\xi - z}. \tag{1}$$

Here $C$ represents any closed contour within the analytic domain of function $f(z)$. This remarkable result shows that the value of an analytic function at an arbitrary point is completely determined by its values at distant points. This mathematical property is independent of any physical assumptions regarding any dynamic equations of motion that may have generated the function $f(z)$.

It will be shown here that the current value of any mathematical function $f(\mathbf{r},t)$ that vanishes at sufficiently large distances and has continuous second partial derivatives is completely determined by its earlier values as given by

$$f(\mathbf{r},t) = -\frac{1}{4\pi} \int d^3\mathbf{r'} \frac{\Box' f(\mathbf{r'},t')}{|\mathbf{r} - \mathbf{r'}|}. \tag{2}$$

The symbol $\Box'$ represents the d'Alembertian operator

$$\Box' f(\mathbf{r'},t') \equiv \nabla'^2 f(\mathbf{r'},t') - \frac{1}{c^2} \frac{\partial^2 f(\mathbf{r'},t')}{\partial t'^2} \tag{3}$$

where $c$ is a constant, such as the speed of light. Here $t'$ corresponds to an earlier (retarded) time given by

$$t' = t - \frac{|\mathbf{r} - \mathbf{r'}|}{c}. \tag{4}$$

From a mathematical point of view, the identity of Eq. (2) can be viewed as a generalization of Green's third identity [2,3].

The function $f(\mathbf{r},t)$ is arbitrary aside from the restrictions mentioned above. This raises an interesting "paradox" as to whether or not the future values of an arbitrary function of this kind are really determined by the past. Unlike a Taylor or Laurent series expansion, Eq. (2) only depends on the second partial derivatives of the function, which need not be analytic. It should be emphasized that the function $f(\mathbf{r},t)$ is not assumed to be a solution to a wave equation or to obey any other dynamic equations of motion.

The remainder of the paper begins with a derivation of Eq. (2) in Section 2. A numerical example is considered in Section 3, which is found to be in excellent agreement with the analytic results. A summary and conclusions are provided in Section 4.

## 2. Mathematical derivation

In order to derive Eq. (2), first consider a function $\phi(\mathbf{r},t)$ that satisfies the wave equation corresponding to the d'Alembertian operator with a source term equal to the function $f(\mathbf{r},t)$ of interest:

$$\nabla^2 \phi(\mathbf{r},t) - \frac{1}{c^2}\frac{\partial^2 \phi(\mathbf{r},t)}{\partial t^2} = -4\pi f(\mathbf{r},t). \tag{5}$$

For example, in classical electromagnetism and in the Lorentz gauge, Eq. (5) would correspond to the retarded scalar potential produced by a charge distribution $f(\mathbf{r},t)$. It is well known [4] that this wave equation has retarded solutions of the form

$$\phi(\mathbf{r},t) = \int d^3\mathbf{r}' \frac{f(\mathbf{r}',t-|\mathbf{r}-\mathbf{r}'|/c)}{|\mathbf{r}-\mathbf{r}'|}. \tag{6}$$

The next step in the derivation of Eq. (2) is to obtain an alternative expression for $\Box\phi$ by taking the appropriate partial derivatives of the right-hand side of Eq. (6), which contains a mild singularity in the integrand at $\mathbf{r}'=\mathbf{r}$. Consider the contribution $I(\varepsilon)$ to the integral from a region of radius $\varepsilon \ll 1$ about the point $\mathbf{r}$, which is on the order of

$$I(\varepsilon) \sim f(\mathbf{r},t)\int_0^\varepsilon 4\pi \frac{\rho^2 d\rho}{\rho} = 2\pi\varepsilon^2 f(\mathbf{r},t). \tag{7}$$

This vanishes in the limit of $\varepsilon \to 0$ and there is no contribution to the integral from the immediate neighborhood of the point $\mathbf{r}$. As a result, we can obtain an arbitrarily close approximation to Eq. (7) by taking

$$\phi(\mathbf{r},t) = \int d^3\mathbf{r}' \frac{f(\mathbf{r}',t-|\mathbf{r}-\mathbf{r}'|/c)}{\left[|\mathbf{r}-\mathbf{r}'|^2 + \varepsilon^2\right]^{1/2}} \tag{8}$$

for sufficiently small values of $\varepsilon$. We will take the limit $\varepsilon \to 0$ at the end of the calculation.

The partial derivatives of $\phi(\mathbf{r},t)$ can be evaluated using the chain rule [5]. It will be convenient to denote the partial derivatives of the function $f(\mathbf{r}',t')$ by

$$\begin{aligned} f_{x'}(\mathbf{r}',t') &\equiv \frac{\partial f(\mathbf{r}',t')}{\partial x'} \\ f_{t'}(\mathbf{r}',t') &\equiv \frac{\partial f(\mathbf{r}',t')}{\partial t'} \end{aligned} \tag{9}$$

with similar notation for the higher partial derivatives. We also define the function $g(\mathbf{r}';\mathbf{r},t) \equiv f(\mathbf{r}',t')$, so that Eq. (8) can be rewritten as

$$\phi(\mathbf{r},t) = \int d^3\mathbf{r}' \frac{g(\mathbf{r}';\mathbf{r},t)}{\left[|\mathbf{r}-\mathbf{r}'|^2 + \varepsilon^2\right]^{1/2}}. \tag{10}$$

Use of the chain rule then gives

$$\frac{\partial g(\mathbf{r}';\mathbf{r},t)}{\partial x} = \frac{\partial f(\mathbf{r}',t')}{\partial t'}\frac{\partial t'}{\partial x} = -\frac{1}{c}f_{t'}(\mathbf{r}',t')\frac{\partial |\mathbf{r}-\mathbf{r}'|}{\partial x} \tag{11}$$

and

$$\frac{\partial g(\mathbf{r}';\mathbf{r},t)}{\partial x'} = \frac{\partial f(\mathbf{r}',t')}{\partial x'} + \frac{\partial f(\mathbf{r}',t')}{\partial t'}\frac{\partial t'}{\partial x'} = f_{x'}(\mathbf{r}',t') - \frac{1}{c}f_{t'}(\mathbf{r}',t')\frac{\partial |\mathbf{r}-\mathbf{r}'|}{\partial x'}. \tag{12}$$

Taking the partial derivative of $\phi(\mathbf{r},t)$ in Eq. (8) with respect to $x$ gives

$$\frac{\partial \phi(\mathbf{r},t)}{\partial x} = \int d^3\mathbf{r}' \{\frac{1}{\left[|\mathbf{r}-\mathbf{r}'|^2 + \varepsilon^2\right]^{1/2}}\frac{\partial f(\mathbf{r}',t-|\mathbf{r}-\mathbf{r}'|/c)}{\partial x} + f(\mathbf{r}',t-|\mathbf{r}-\mathbf{r}'|/c)\frac{\partial}{\partial x}\frac{1}{\left[|\mathbf{r}-\mathbf{r}'|^2 + \varepsilon^2\right]^{1/2}}\}. \tag{13}$$

Using the results of the chain rule in Eq. (11) gives

$$\frac{\partial \phi(\mathbf{r},t)}{\partial x} = \int d^3\mathbf{r}' \{-\frac{f_{t'}(\mathbf{r}',t-|\mathbf{r}-\mathbf{r}'|/c)}{\left[|\mathbf{r}-\mathbf{r}'|^2 + \varepsilon^2\right]^{1/2}}\frac{1}{c}\frac{\partial |\mathbf{r}-\mathbf{r}'|}{\partial x} - f(\mathbf{r}',t-|\mathbf{r}-\mathbf{r}'|/c)\frac{\partial}{\partial x'}\frac{1}{\left[|\mathbf{r}-\mathbf{r}'|^2 + \varepsilon^2\right]^{1/2}}\}. \tag{14}$$

Here the sign of the second term in the integral has been reversed and the partial derivative there is now with respect to $x'$, which makes use of the fact that

$$\partial |\mathbf{r}-\mathbf{r}'|/\partial x = -\partial |\mathbf{r}-\mathbf{r}'|/\partial x'. \tag{15}$$

The second term in the integral of Eq. (14) can be rewritten using integration by parts [6] with respect to $x'$, which gives

$$\frac{\partial \phi(\mathbf{r},t)}{\partial x} = \int d^3\mathbf{r}' \{-\frac{f_{t'}(\mathbf{r}',t-|\mathbf{r}-\mathbf{r}'|/c)}{\left[|\mathbf{r}-\mathbf{r}'|^2 + \varepsilon^2\right]^{1/2}}\frac{1}{c}\frac{\partial |\mathbf{r}-\mathbf{r}'|}{\partial x} + \frac{1}{\left[|\mathbf{r}-\mathbf{r}'|^2 + \varepsilon^2\right]^{1/2}}\frac{\partial f(\mathbf{r}',t-|\mathbf{r}-\mathbf{r}'|/c)}{\partial x'}\}. \tag{16}$$

Here we have made use of the fact that $f(\mathbf{r},t)$ has been assumed to vanish at large distances. To be more precise, Eq. (16) neglects the integral $I_B$ given by

$$I_B = -\int_{-\infty}^{\infty} dy' dz' \frac{f(\mathbf{r}',t')}{|\mathbf{r}'-\mathbf{r}|}, \tag{17}$$

where the limit of $x' \to \pm\infty$ is to be taken. The integral of Eq. (17) will be negligible if $|f(\mathbf{r}',t')|$ drops off faster than $1/|\mathbf{r}'-\mathbf{r}|$ at large distances. (For example, $I_B = 0$ if $|f(\mathbf{r}',t')| \leq \alpha/|\mathbf{r}'-\mathbf{r}|^2$ in the limit of large $|\mathbf{r}'-\mathbf{r}|$, where $\alpha$ is any finite constant.) Integration by parts also requires [6] that

$f(\mathbf{r}',t')$ have continuous first partial derivatives with respect to $x$ ($C^1$). A second integration by parts (below) will require that it have continuous second partial derivatives.

Eq. (16) can be further simplified by using the chain rule to perform the differentiation with respect to $x'$. From Eq. (12) this gives

$$\frac{\partial \phi(\mathbf{r},t)}{\partial x} = \int d^3\mathbf{r}' \{ -\frac{f_{t'}(\mathbf{r}',t-|\mathbf{r}-\mathbf{r}'|/c)}{\left[|\mathbf{r}-\mathbf{r}'|^2 + \varepsilon^2\right]^{1/2}} \frac{1}{c} \frac{\partial |\mathbf{r}-\mathbf{r}'|}{\partial x}$$
$$-\frac{f_{t'}(\mathbf{r}',t-|\mathbf{r}-\mathbf{r}'|/c)}{\left[|\mathbf{r}-\mathbf{r}'|^2 + \varepsilon^2\right]^{1/2}} \frac{1}{c} \frac{\partial |\mathbf{r}-\mathbf{r}'|}{\partial x'} + \frac{f_{x'}(\mathbf{r}',t-|\mathbf{r}-\mathbf{r}'|/c)}{\left[|\mathbf{r}-\mathbf{r}'|^2 + \varepsilon^2\right]^{1/2}} \}. \tag{18}$$

It can be seen using Eq. (15) once again that the first and second terms in the integral cancel out, which gives the simplified result

$$\frac{\partial \phi(\mathbf{r},t)}{\partial x} = \int d^3\mathbf{r}' \frac{f_{x'}(\mathbf{r}',t-|\mathbf{r}-\mathbf{r}'|/c)}{\left[|\mathbf{r}-\mathbf{r}'|^2 + \varepsilon^2\right]^{1/2}}. \tag{19}$$

This process can be repeated in the same way to show that

$$\frac{\partial^2 \phi(\mathbf{r},t)}{\partial x^2} = \int d^3\mathbf{r}' \frac{f_{x'x'}(\mathbf{r}',t-|\mathbf{r}-\mathbf{r}'|/c)}{\left[|\mathbf{r}-\mathbf{r}'|^2 + \varepsilon^2\right]^{1/2}} \tag{20}$$

where $f_{x'x'}$ denotes the second partial derivative of $f$ with respect to $x'$ as in Eq. (9). Here the integration by parts is valid if $|f_{x'}(\mathbf{r}',t')|$ drops off faster than $1/|\mathbf{r}'-\mathbf{r}|$ (as would be the case if $|f_{x'}(\mathbf{r}',t')| \le \beta / |\mathbf{r}'-\mathbf{r}|^2$ in the limit of large $|\mathbf{r}'-\mathbf{r}|$ where $\beta$ is any finite constant, for example). Integration by parts also requires [6] that both functions be continuously differentiable ($C^1$). Since it is $f_{x'}(\mathbf{r}',t')$ that appears here in the integration by parts, this requires that $f(\mathbf{r}',t')$ have continuous second partial derivatives with respect to $x'$ ($C^2$).

The partial derivatives with respect to $y$ and $z$ can also be evaluated in the same way to give

$$\left(\nabla^2 - \frac{1}{c^2}\frac{\partial^2}{\partial t^2}\right)\phi(\mathbf{r},t) = \int d^3\mathbf{r}' \frac{\left(f_{x'x'} + f_{y'y'} + f_{z'z'} - f_{t't'}/c^2\right)}{\left[|\mathbf{r}-\mathbf{r}'|^2 + \varepsilon^2\right]^{1/2}} = -4\pi f(r,t). \tag{21}$$

Eq. (21) also includes the second partial derivative of Eq. (8) with respect to time, which is straightforward. The arguments of the functions have been omitted here in order to shorten the equation.

The right-hand side of Eq. (21) follows from the fact that $\phi(\mathbf{r},t)$ is arbitrarily close to a solution to Eq. (5), the wave equation. Solving for $f(\mathbf{r},t)$ and taking the limit of $\varepsilon \to 0$ gives Eq. (2), as desired. This derivation is somewhat similar to that of Green's identities [2,3], which are also based on integration by parts. In fact, Eq. (2) reduces to Green's third identity with these boundary conditions in the static limit where the function $f(\mathbf{r},t)$ is independent of time.

The proof given above assumes that the partial derivatives can be taken inside the integral, which is often taken for granted. This is equivalent to interchanging the order of two limits, which is valid provided that the corresponding sequences are uniformly convergent [7]. That should be the case here since the function has already been required to have continuous second partial derivatives. The fact that Eq. (2) reduces to one of Green's identities suggests that the convergence properties here are the same as in the proof by Green, which was also based on functions with continuous second partial derivatives.

The derivation of Eq. (2) does not depend on the value of the constant c. As a result, we could define a new operator $\Box_{c'}$ by replacing the constant $c$ with $c'$ in Eq. (3), in which case Eq. (2) will still hold using $\Box_{c'}$. The value of $c'$ can also be negative, which would correspond to the use of the advanced propagator in Eq. (6) instead of the usual retarded propagator. Both choices are possible, but the use of the retarded propagator gives the results of interest here; this choice is somewhat analogous to the usual use of retarded propagators based on the initial conditions. Although the results described here were derived using the Green's function associated with the d'Alembertian operator, it seems likely that similar results could also be obtained using other types of wave equations.

Eq. (2) can be written in a covariant form as

$$f(\mathbf{x}) = -\frac{1}{2\pi} \int_{-\infty}^{\infty} d^4\mathbf{x}' \theta(\mathbf{x}-\mathbf{x}') \Box' f(\mathbf{x}') \delta[(x_\mu - x'_\mu)(x^\mu - x'^\mu)]. \tag{22}$$

Here $\mathbf{x}$ is the usual 4-vector with components $(ct,\mathbf{r})$ and $\theta(\mathbf{x})$ is defined [8] as being equal to 1 if $x^0 > 0$, ½ if $x^0 = 0$, and 0 otherwise. Eq. (22) can be shown to be equivalent to Eq. (2) by rewriting it in the form

$$f(\mathbf{x}) = -\frac{1}{2\pi} \int_{-\infty}^{\infty} d^3\mathbf{r}' \int_{-\infty}^{t} cdt' \Box' f(\mathbf{x}') \delta\left[\left(|\mathbf{r}-\mathbf{r}'|+c(t-t')\right)\left(|\mathbf{r}-\mathbf{r}'|-c(t-t')\right)\right]. \tag{23}$$

The integral in Eq. (23) can be evaluated by making a change of variables from $t'$ to a new variable $\chi$ defined by

$$\chi = \left(|\mathbf{r}-\mathbf{r}'|+c(t-t')\right)\left(|\mathbf{r}-\mathbf{r}'|-c(t-t')\right). \tag{24}$$

With this change of variables Eq. (24) becomes

$$f(\mathbf{x}) = -\frac{1}{2\pi} \int_{-\infty}^{\infty} d^3\mathbf{r}' \int_{-\infty}^{|\mathbf{r}-\mathbf{r}'|^2} cd\chi \frac{\Box' f(\mathbf{x}') \delta[\chi]}{c\left(|\mathbf{r}-\mathbf{r}'|+c(t-t')\right)+c\left(|\mathbf{r}-\mathbf{r}'|-c(t-t')\right)}. \tag{25}$$

where $t'$ is now a function of $\chi$. The second term in the denominator of Eq. (25) vanishes when the argument of the $\delta$-function is zero at the retarded time while the remaining term in the denominator gives the $1/|\mathbf{r}-\mathbf{r}'|$ dependence of Eq. (2), which is thus equivalent to Eq. (22). The covariant form of Eq. (22) may be useful in quantum field theory, which was the original motivation [9] for my interest in this topic.

Eq. (2) raises an interesting question as to whether or not the future values of an arbitrary function of this kind are really determined by its past values, and whether we could use this property to predict future variables of interest, such as the weather or the stock market. It can be seen from Eq. (7) that the contribution to $f(\mathbf{r},t)$ from the time interval $\Delta t$ just before time $t$ becomes vanishingly small as $\Delta t \to 0$. This shows that $f(\mathbf{r},t)$ really is determined entirely by its past values. Suppose that we attempt to predict the value of the function at time $t$ using only its values up to time $t - \Delta t$. That is equivalent to ignoring a relatively small part of the total integral in Eq. (2), and the prediction should be increasingly accurate over shorter time intervals $\Delta t$.

Nevertheless, the function is still arbitrary in the sense that its value could change by an arbitrarily large amount in the final time interval $\Delta t$ over which the prediction is made. If that turns out to be the case, it would simply correspond to an anomalously large value of $\Box f(\mathbf{r},t)$ over the final time interval. Any predictive power of Eq. (2) is due to the fact that such an anomaly is unlikely, and there is no real paradox. This technique is unlikely to have any practical advantages over other predictive methods that are based on physical models of the system of interest [10]; what is of interest here is that there is no underlying model at all.

Eqs. (2) and (21) were derived using the retarded solution to a specific wave equation, namely that of Eq. (5). But it should be emphasized that these results are valid for any function that vanishes at infinity and has continuous second partial derivatives. The point is that Eq. (2) is more general than any wave equation, since it does not assume any dynamic equations of motion. Other integral identities can probably be derived using other wave equations, as mentioned above, in which case those identities would also be valid for arbitrary functions of this kind. As previously noted, a similar identity can also be derived using the advanced Green's function.

## 3. Numerical example

The integral identity of Eq. (2) is sufficiently surprising that it may be worthwhile to compare its predictions with the results of a numerical calculation based on a specific example. The numerical example also provides a graphic illustration of the contribution to the integral of Eq. (2) from different locations and retarded times.

Let $f_0(\mathbf{r})$ be a static function given by

$$f_0(\mathbf{r}) = \begin{cases} \cos^4(r) & r \leq \pi/2 \\ 0 & r > \pi/2. \end{cases} \quad (26)$$

We can then construct a time-dependent function that corresponds to $f_0(\mathbf{r})$ translated along the z axis at a constant velocity $v$:

$$f(\mathbf{r},t) = f_0(x, y, z - vt). \quad (27)$$

A plot of $f(\mathbf{r},t)$ in the $x$-$z$ plane (y=0) is shown in Fig. 1 at time $t = 0$, when the function is centered at the origin.

Fig. 2 shows a plot of the retarded function $f(\mathbf{r}',t - |\mathbf{r}-\mathbf{r}'|/c)$ for the case of $\mathbf{r} = 0$, $t = 0$ and $v = c/2$. It can be seen that the retarded function is compressed in the direction of travel and stretched out in the opposite direction due to the fact that the retarded contribution has to "catch up" with the moving distribution at the appropriate retarded time.

The d'Alembertian of this function in the $x$-$z$ plane is shown in Fig. 3, also at time $t = 0$. Fig. 4 shows a plot of the retarded d'Alembertian as a function of $x'$ and $z'$ at $\mathbf{r} = 0$ and $t = 0$. This is the function that would be integrated to calculate the value of $f(\mathbf{r},t)$ at the origin and at $t = 0$ when using Eq. (2).

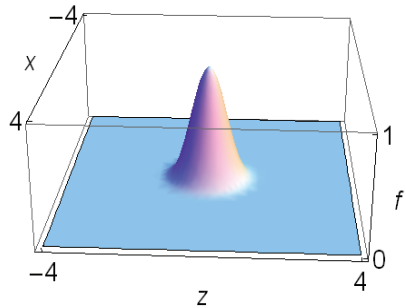

Fig. 1. A plot of the sample function $f(\mathbf{r},t)$ from Eq. (27) in the $x$-$z$ plane at time $t = 0$ (arbitrary units).

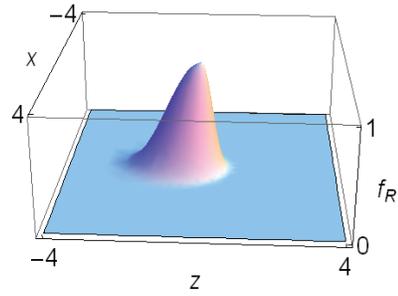

Fig. 2.  A plot of the retarded sample function $f(\mathbf{r}',t-|\mathbf{r}-\mathbf{r}'|/c)$ in the $x'$-$z'$ plane at $\mathbf{r}=0$ and $t=0$ (arbitrary units).

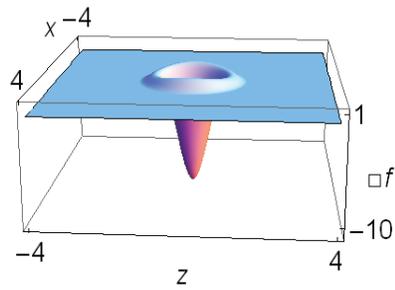

Fig. 3.  A plot of the d'Alembertian $\Box f$ of the sample function in the $x$-$z$ plane at time $t=0$ (arbitrary units).

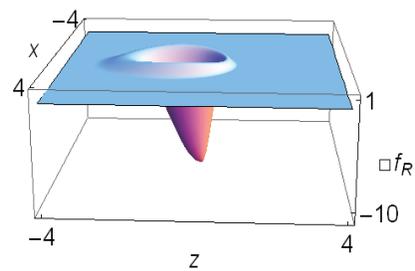

Fig. 4.  A plot of the retarded d'Alembertian $\Box f_R$ of the sample function in the $x'$-$z'$ plane for $\mathbf{r}=0$ and $t=0$ (arbitrary units).

Figure 5 shows a plot of the value $f_{calc}(\mathbf{r},t)$ of the function calculated using Eq. (2).  For comparison purposes, the expected value from Eq. (27) is plotted as a dashed line.  These results were obtained by numerical integration using Mathematica.  Similar results obtained for a variety of other conditions and functions were all in agreement with the expected results to within the numerical precision (typically 6 significant digits).  These numerical calculations show that Eq. (2) does give the correct results for $f(\mathbf{r},t)$ at least for the examples that were considered, and they illustrate the way in which the retarded time affects the contribution to the integral from different locations in space.

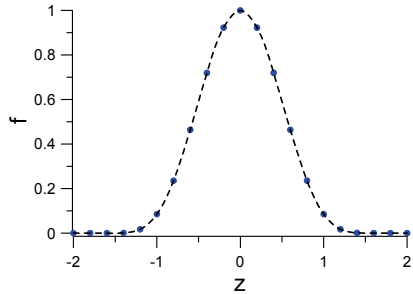

Fig. 5. A plot of the value of the function $f(\mathbf{r},t)$ calculated from Eq. (2) (blue dots) as a function of $z$ compared with the correct value from Eq. (27) (dashed line). These results correspond to $x = y = t = 0$, but similar agreement was obtained for other values of $\mathbf{r}$. (Arbitrary units).

## 4. Summary and conclusions

A new integral identity shows that the current value of any mathematical function $f(\mathbf{r},t)$ is completely determined by its prior values at all other locations $\mathbf{r'}$ and retarded times $t' \leq t$, provided that it has continuous second partial derivatives and that the function and its first partial derivatives decrease faster than $1/|\mathbf{r}-\mathbf{r'}|$ at sufficiently large distances. It should be emphasized once again that the function $f(\mathbf{r},t)$ was not assumed to be a solution to a wave equation or to obey any other dynamic equations of motion. This integral identity was derived using Green's function techniques and it can be viewed as a generalization of Green's third identity, which has applications in applied physics [3]. It is somewhat similar to Cauchy's theorem for analytic functions, but here $f(\mathbf{r},t)$ need not be analytic.

Eqs. (2) and (22) show that functions of this kind are constrained in a way that may seem somewhat surprising. The function $f(\mathbf{r},t)$ is arbitrary (aside from the above requirements) and yet its current value is determined by its prior values. As discussed in section 2, the value of $f(\mathbf{r},t)$ can change by an arbitrarily large amount during a small time interval $\Delta t$ provided that the partial derivatives are correspondingly large, and there is no contradiction with the fact that the function is arbitrary. A similar identity can also be derived using advanced instead of retarded Green's functions, and these results do not imply any preferred direction for the "arrow of time".

Functions of this kind occur in many fields of physics, such as classical electromagnetism and acoustics. Other forms of integral identities have practical applications, and that may also be the case for the identities of Eqs. (2) and (22).

I would like to acknowledge stimulating discussions with Alain Aspect, Nicolas Gisin, Alex Kaplan, Anthony Leggett, and Todd Pittman. This work was supported in part by the National Science Foundation (NSF) under grant 0652560.

## Appendix

It was suggested by an anonymous referee that there is a counter-example to Eq. (2) in which the function $f(\mathbf{r},t)$ includes a solution to the homogeneous wave equation. It will be shown here that Eq. (2) correctly describes functions of that kind if the field was generated by a source at some time in the past. It will also be shown that the nonphysical case in which there is no source and the wave equation is homogeneous at all times does not satisfy the conditions assumed in the derivation of Eq. (2) and therefore does not constitute a counter-example.

Consider an example in which the function $f(\mathbf{r},t)$ of interest has the form

$$f(\mathbf{r},t) = f_i(\mathbf{r},t) + f_h(\mathbf{r},t). \tag{A1}$$

Here $f_i(\mathbf{r},t)$ is a solution to the inhomogeneous wave equation with a source term $S_i(\mathbf{r},t)$

$$\nabla^2 f_i(\mathbf{r},t) - \frac{1}{c^2}\frac{\partial^2 f_i(\mathbf{r},t)}{\partial t^2} = -4\pi S(\mathbf{r},t) \tag{A2}$$

while $f_h(\mathbf{r},t)$ is a solution to the homogenous equation with $S=0$. Suppose that $f_i(\mathbf{r},t)$ is correctly given by Eq. (2). Then adding the inhomogeneous solution $f_h(\mathbf{r},t)$ will change the left-hand side of Eq. (2) but have no effect on the right-hand side of that equation because $\Box' f_h(\mathbf{r}',t') = 0$. It was suggested that this provides a counter-example to the derivation of Eq. (2).

The error in this argument can be illustrated by considering two different examples in which the wave equation is homogeneous at the time t when Eq. (2) is to be applied. For simplicity, we will consider spherically-symmetric waves as illustrated in Fig. 6. In the first example, it will be assumed that the field was generated by a spherically-symmetric source that was nonzero only for a small time interval about an earlier time $t' \ll t$ and turned off before the time of interest. The source is assumed to have been localized at a distance R from the origin at that time, as illustrated in Fig. 6a. This produces a spherical wave propagating towards the origin, as indicated by the red arrow, as well as a wave propagating outwards (not shown in the figure). In this case the wave equation is not homogenous at the relevant retarded times and $\Box' f(\mathbf{r}',t') \neq 0$ in the integral of Eq. (2). As a result, the correct results are obtained even though the wave equation is homogeneous at the time of interest.

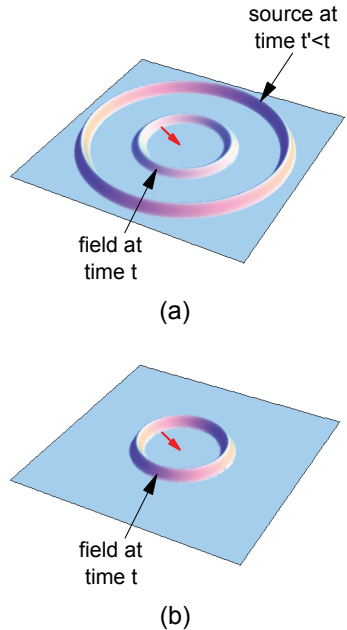

Fig. 6. Two examples of solutions to a wave equation that is homogeneous at the time t of interest. (a) A spherically-symmetric source at time $t' \ll t$ generates a spherical wave that propagates towards the origin. The value of the field in the x-y plane is shown. Eq. (2) correctly describes this situation. (b) A spherical wave is assumed to propagate toward the origin from infinity with no source, in which case the wave equation is homogeneous at all times. This nonphysical situation does not satisfy the assumptions made in the derivation of Eq. (2) and it does not constitute a counter-example.

10The other example of interest is illustrated in Fig. 6b, where it is assumed that the wave equation is homogenous at all times. In that case a spherical wave propagates inwards from infinity with no source, which is mathematically equivalent to taking the limit $R \to \infty$. In this case it is indeed true that the left-hand side of Eq. (2) would be nonzero while the right hand side would be zero because $\Box' f(\mathbf{r}',t') = 0$ at all times. But in this example the field drops off as $1/|\mathbf{r}'-\mathbf{r}|$ at the relevant retarded times, whereas the derivation of Eq. (2) explicitly assumed that the function drops off faster than that for all times $t' \leq t$. This does not constitute a counter-example to Eq. (2) since the conditions assumed in its proof are not satisfied in this case; Eq. (2) simply does not apply to situations of this kind.

From a physical point of view, it seems reasonable to suppose that any field must have been generated by a source at some time in the past as in Fig. 6a, in which case $f(\mathbf{r}',t')$ and its partial derivatives would be zero beyond some finite distance from the origin and Eq. (2) would apply. In any event, solutions to the homogeneous wave equation do not provide a counter-example to the proof of Eq. (2) as stated. It should also be emphasized once again that Eq. (2) is true in general and it is not limited to solutions to the wave equation; the fields in Fig 6 are merely examples.

**References**

1. E. Butkov, *Mathematical Physics* (Addison-Wesley, Reading, Mass., 1968).
2. W. Strauss, *Partial Differential Equations: An Introduction* (Wiley, New York, 2008).
3. R.J. Blakely, *Potential Theory in Gravity and Magnetic Applications* (Cambridge U. Press, Cambridge, 1996).
4. J.D. Jackson, *Classical Electrodynamics* (Wiley, New York, 1962).
5. A.E. Taylor, *Calculus with Analytic Geometry* (Prentice-Hall, Englewood Cliffs, N.J., 1966).
6. Evans, L.C., *Partial Differential Equations* (American Mathematical Society, Providence, R.I., 1998).
7. W. Kaplan, *Advanced Calculus* (Addison-Wesley, Reading, Mass., 1952).
8. J.M. Jauch and F. Rohrlich, *The Theory of Photons and Electrons*, 2nd ed. (Springer-Verlag, Berlin, 1980).
9. J.D. Franson, Phys. Rev. A **84**, 033809 (2011).
10. A. Gelb, ed., *Applied Optimal Estimation* (MIT Press, Cambridge, Mass., 1974).